\documentclass[showpacs,amsmath,amssymb,aps,10pt,reprint,superscriptaddress,prb]{revtex4-1}
\usepackage{graphicx}
\usepackage{bm}
\usepackage[breaklinks=true,colorlinks=true,linkcolor=blue,urlcolor=blue,citecolor=blue]{hyperref}
\usepackage{graphics}
\usepackage{dcolumn}
\usepackage{amsmath,amssymb}
\usepackage{mathdots}
\usepackage{natbib}
\usepackage{soul,color}
\usepackage{epstopdf}
\usepackage[note-name]{notes2bib}
\begin{document}

\title{Zero-bias Shapiro steps in asymmetric pinning nanolandscapes}

\author{O.~V.~Dobrovolskiy}
    %\email[Corresponding author: ]{Dobrovolskiy@Physik.uni-frankfurt.de}
    \affiliation{Physikalisches Institut, Goethe University, 60438 Frankfurt am Main, Germany}
    \affiliation{Physics Department, V. Karazin Kharkiv National University, 61077 Kharkiv, Ukraine}
\author{V. V. Sosedkin}
    \affiliation{Physics Department, V. Karazin Kharkiv National University, 61077 Kharkiv, Ukraine}
\author{R. Sachser}
    \affiliation{Physikalisches Institut, Goethe University, 60438 Frankfurt am Main, Germany}
\author{V.~A.~Shklovskij}
\author{R. V. Vovk}
    \affiliation{Physics Department, V. Karazin Kharkiv National University, 61077 Kharkiv, Ukraine}
\author{M. Huth}
    \affiliation{Physikalisches Institut, Goethe University, 60438 Frankfurt am Main, Germany}

\begin{abstract}
The coherent nonlinear dynamics of Abrikosov vortices in asymmetric pinning nanolandscapes is studied by theoretical modeling and combined microwave and dc electrical resistance measurements. The problem is considered on the basis of a single-vortex Langevin equation within the framework of a stochastic model of anisotropic pinning. When the distance over which Abrikosov vortices are driven during one half ac cycle coincides with one or a multiple of the nanostructure period, Shapiro steps appear in the current-voltage curves (CVCs) as a general feature of systems whose evolution in time can be described in terms of a particle moving in a periodic potential under combined dc and ac stimuli. While a dc voltage appears in response to the ac drive, the addition of a dc bias allows one to diminish the rectified voltage and eventually to change its sign when the extrinsic dc bias-induced asymmetry of the pinning potential starts to dominate the intrinsic one. This rectified negative voltage in the CVCs becomes apparent as \emph{zero-bias} Shapiro steps, which are theoretically predicted and experimentally observed for the first time.
\end{abstract}
\maketitle

\section{Introduction}
Periodic arrays of pinning sites are known to be effective for reducing the dissipation by Abrikosov vortices in type-II superconductors. In the presence of periodic pinning structures, the dynamics of vortices can be described as their motion in some periodic pinning potential \cite{Bra95rpp}. In the most simple case such a potential is constant in one direction and periodic in the other, i.\,e. it is a pinning potential of the \emph{washboard} type \cite{Dob17pcs}. This type of pinning potential allows one to derive \emph{analytical} expressions for the experimentally measurable quantities (dc voltage and absorbed ac power) for \emph{arbitrary} values of the driving parameters (dc bias, ac amplitude and frequency) and temperature.

An intriguing effect in the vortex dynamics appears when the pinning potential is asymmetric~\cite{Lee99nat,Vil03sci,Sil06nat,Zap96prl,Ust04prl,Miz06jap}. In this case the reflection symmetry of the pinning force is broken and thus, the depinning currents measured under current reversal are not equal. This is known as the \emph{ratchet} effect, which in the context of the vortex dynamics is associated with systems \cite{Plo09tas} where the vortex can acquire a motion whose direction is determined only by the pinning potential asymmetry with respect to the current direction reversal. In general, depending on the way to bring asymmetry into a system, one can distinguish between a fixed \emph{intrinsic} asymmetry caused by the spatial asymmetry of the potential, and a tunable \emph{extrinsic} asymmetry invoked by a dc bias. A vortex ratchet with internal pinning asymmetry is called a \emph{rocking ratchet}, while a vortex ratchet with external pinning asymmetry is known as a tilted-potential or \emph{tilting ratchet}. A tilting ratchet was theoretically considered in \cite{Shk11prb} for a cosine washboard pinning potential (WPP) where its tilt is caused by a dc bias. In Ref. \cite{Shk11prb} exact expressions for the experimentally accessible quantities (electrical voltage and absorbed ac power) were derived by using the matrix continued fractions technique \cite{Shk08prb}. This method was recently extended \cite{Shk14pcm} to an asymmetric potential of the washboard type. In this work, the theoretical results of Ref. \cite{Shk14pcm} will be used for analysing the current-voltage curves (CVC) of superconductors with ratchet WPPs.

Another important effect consists in switching between the direct and the reversed net motion of the vortices and is termed the \emph{ratchet reversal} effect \cite{Shk14pcm}. This effect manifests itself as a sign change in the rectified voltage as a function of the driving force. Though a dozen of mechanisms for ratchet reversals are discussed in literature, as briefly outlined in the introductory part of Ref. \cite{Shk14pcm}, one of the simplest mechanisms relies upon \emph{competition} of the intrinsic and the extrinsic asymmetries of the pinning potential. It is this mechanism which will be discussed in this work for the vortex dynamics under combined dc and ac drives in a ratchet WPP.

On the experimental side, our previous work was concerned with investigations of the guided vortex motion in Nb films \cite{Dob12tsf} with symmetric nanogrooves \cite{Dob12njp}, the interplay of vortex guiding with the Hall effect \cite{Dob16sst} and the dynamics of vortices at microwave frequencies \cite{Dob15mst,Dob15snm,Dob15apl,Dob15met,Sil17inb}. On the one hand, Shapiro steps were revealed in the CVCs of Nb films with periodically arranged symmetric nanogrooves \cite{Dob15mst,Dob15snm}. That provided evidence that when the location of vortices geometrically matches the location of the pinning sites, the vortex dynamics is coherent, so that the dynamics of the whole vortex ensemble can be analyzed in terms of the single-vortex dynamics \cite{Luq07prb}. In general, Shapiro steps appear in the current-voltage curves (CVCs) as a general feature of systems whose evolution in time can be described in terms of a particle moving in a periodic potential under combined dc and ac stimuli. The observation of Shapiro \cite{Sha63prl} steps in different systems dates back to the works of Fiory \cite{Fio71prl,Fio73prb}, Martinoli \cite{Mar75ssc,Mar76prl} and others \cite{Day67prv,Ben90prl,Loo99prb,Mat09prb,Siv03prl,Naw13prl,Rei00prb,Rei15prbS}.

On the other hand, the dc current polarity has been revealed to be crucial for the reduction of the so-called \emph{depinning} frequency \cite{Git66prl,Cof91prl,Pom08prb,Sil17inb} in Nb films with asymmetric grooves, thus allowing for the design of microwave cutoff filters \cite{Dob15apl} and fluxonic metamaterials \cite{Dob15met}. In addition, an analysis of experimental data on the microwave power absorption using the approach outlined in Refs. \cite{Shk08mmt,Shk12inb,Shk13ltp} has allowed us to deduce the coordinate dependences of the pinning potentials in these samples \cite{Dob15vor}. Since the deduced coordinate dependences turned out asymmetric \cite{Dob15vor}, this motivated us to theoretically analyze the CVCs for the films with asymmetric WPPs and to examine the predicted effects experimentally.

Here, we theoretically analyze the appearance of \emph{zero-bias} Shapiro steps in the CVCs of superconductors with asymmetric washboard pinning landscapes as functions of the asymmetry strength parameter and the ac amplitude. On the experimental side, studying the voltage response by combined microwave ($1$\,MHz -- $1$\,GHz) and dc electrical resistance measurements, we reveal the theoretically predicted zero-bias Shapiro steps which are absent in the CVCs of superconducting films with symmetric WPPs. These zero-bias Shapiro steps originate from the negative ratchet effect which is caused by the different groove slopes' steepnesses in the studied Nb films.

\section{Theoretical model}
\label{sTheor}
\begin{figure}
    \centering
    \includegraphics[width=0.42\textwidth]{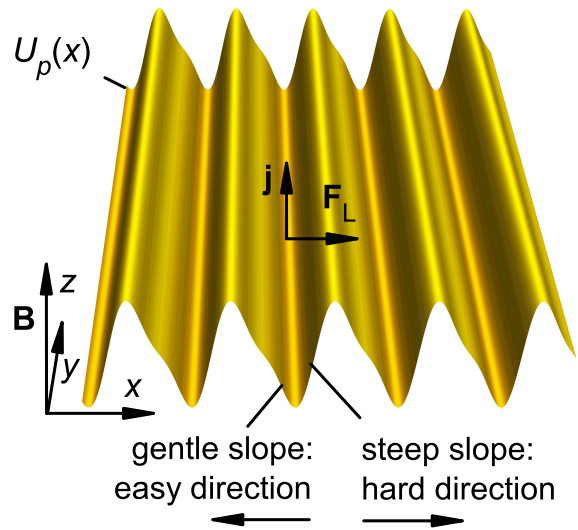}
    \caption{The channels of the ratchet WPP $U_p(x)$ computed by Eq. \eqref{eWPP} with $e = 0.5$ are parallel to the $y$-axis along which the transport current density vector $\mathbf{j}=\mathbf{j}^{dc}+\mathbf{j}^{ac}\cos\omega t$ is applied. $\mathbf{B}$ is the magnetic field and $\mathbf{F}_L$ is the Lorenz force for a vortex. Indicated are the steep-slope and gentle-slope directions of the pinning potential along the $x$ axis.}
   \label{f1}
\end{figure}

\begin{figure}
    \centering
    \includegraphics[width=0.9\linewidth]{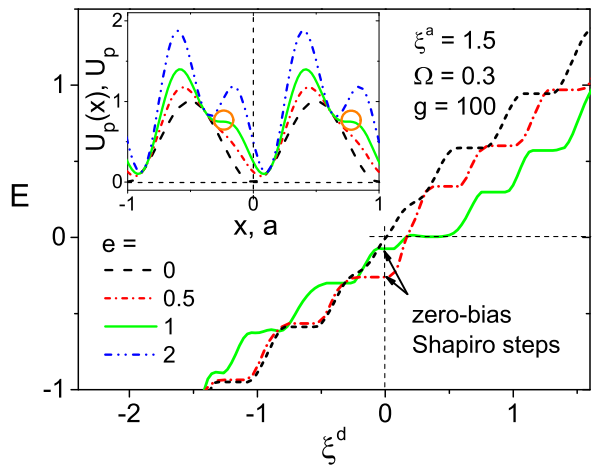}
    \caption{$E(\xi^d)$ by Eq.~\eqref{eCVC} for a series of values of the asymmetry parameter $e$. Inset: WPPs calculated by Eq.~\eqref{eWPP} for the same series of $e$. The cosine WPP ($e=0$) was studied in Refs. \cite{Shk08prb,Shk11prb}. The ratchet WPP with $e=0.5$ is exemplary for the experimental sample in Sec. \ref{sExp}. The circles mark saddle points in the WPP for $e=1$. Also a double-well ratchet WPP, as representative for $e=2$, can be accounted for \cite{Shk14pcm}, but it is not considered here.}
   \label{f2}
\end{figure}

The geometry of the problem is sketched in Fig.~\ref{f1}. Its theoretical treatment relies upon the Langevin equation for a vortex moving with velocity $\mathbf{v}$ in a magnetic field $\mathbf{B}=\mathbf{n}B$, where $B\equiv|\mathbf{B}|$, $\mathbf{n}=n\mathbf{z}$, $\mathbf{z}$~is the unit vector in the $z$ direction and $n=\pm 1$, which in neglect of the Hall effect reads \cite{Shk08prb}
\begin{equation}
        \label{eLE}
        \eta\mathbf{v} = \mathbf{F}_{L}+\mathbf{F}_{p}+\mathbf{F}_{th},
\end{equation}
where $\mathbf{F}_{L}=n(\Phi_{0}/c)\mathbf{j}\times\mathbf{z}$ is the Lorentz force, $\Phi_{0}$ is the magnetic flux quantum, and $c$ is the speed of light. In Eq. \eqref{eLE} $\mathbf{j}=\mathbf{j}(t)= \mathbf{j}^{dc}+ \mathbf{j}^{ac} \cos\omega t$, where $\mathbf{j}^{dc}$ and $\mathbf{j}^{ac}$ are the dc and ac current density amplitudes and $\omega$ is the angular frequency. $\mathbf{F}_{p}=-\nabla U_p(x)$ is the anisotropic pinning force, where $U_p(x)$ is a ratchet WPP. $\mathbf{F}_{th}$ is the thermal fluctuation force represented by Gaussian white noise and $\eta$ is the vortex viscosity.
\quad The ratchet WPP is modeled by
\begin{equation}
        \label{eWPP}
        U_p(x) = (U_p/2) [1-\cos kx + e(1 -\sin 2kx)/2],
\end{equation}
where $k=2\pi/a$. Here $a$ is the period and $U_p$ is the depth of the WPP. In Eq. \eqref{eWPP} $e$ is the asymmetry parameter allowing for tuning the asymmetry strength. In particular, Eq. \eqref{eWPP} yields a symmetric WPP when $e =0$ and a double-well WPP when $e = 2$, refer to the inset of Fig. \ref{f2}. In what follows, we will focus on the case $e=0.5$, as representative for the most commonly~\cite{Bar94epl,Han96inc,Zap96prl,Mat00prl,Pop00prl,Zar09pre,Arz11prl} used ratchet potential. Evidently, the left and right barriers of the ratchet WPP have different steepnesses. For definiteness, the positive current polarity corresponds to the vortex motion in the positive direction of the $x$-axis direction against the steep slope of the WPP in Fig. \ref{f1}.

The Langevin equation \eqref{eLE} can be solved in terms of the matrix continued fractions \cite{Shk14pcm}. For this one introduces dimensionless parameters, namely, the dc density $\xi^d = j^d/j_c$ and the ac density amplitude $\xi^a = j^a/j_c$. Here $j^{d,a} = |j^{dc,ac}|$ and $j_c = c U_p k /2\Phi_0$ is the depinning current density corresponding to the WPP barrier vanishment. Further dimensionless parameters are the frequency $\Omega = \omega/\omega_p$, where $\omega_p\equiv 1/\hat\tau = U_p k^2/2\eta$ is the depinning frequency with $\hat\tau $ being the relaxation time for the vortex, the coordinate $\textsl{x} = kx$ and the inverse temperature $g = 2U_p/T$. The experimentally deducible quantities are the microwave power absorbed by vortices and the dc electric field strength. It is the dc electric field strength on which we focus in this work. Namely, in units of the flux-flow resistivity, $\rho_f = B\Phi_0/\eta c^2$, the time-independent average dc electric field $E$ reads \cite{Shk14pcm}
\begin{equation}
    \label{eCVC}
    E= \nu(\xi^d, \xi^a, \Omega, g)\xi^d.
\end{equation}
Here $\nu$ is the $(\xi^d, \xi^a, \Omega, g)$-dependent effective \emph{nonlinear} mobility of the vortex under the influence of the dimensionless generalized moving force in the $x$ direction and it is expressed in terms of the matrix continued fractions \cite{Shk14pcm}.

In what follows we discuss the modification of the CVCs $E(\xi^d)$ calculated by Eq. \eqref{eCVC} for a series of values of the asymmetry parameter $e$ and the ac amplitude $\xi^a$ at the reduced temperature $g = 2U_p/T = 100$. Note, this corresponds to the experimental situation, e.\,g. Nb thin films with nanogrooves, where the pinning activation energy is of the order of $5000$\,K \cite{Dob12njp} and the measurements are conducted in the vicinity of the superconducting transition temperature $T \simeq T_c \sim 8$\,K. The computing procedure with the matrix continued fractions is detailed in Ref. \cite{Shk14pcm}.

\begin{figure}
    \centering
    \includegraphics[width=0.9\linewidth]{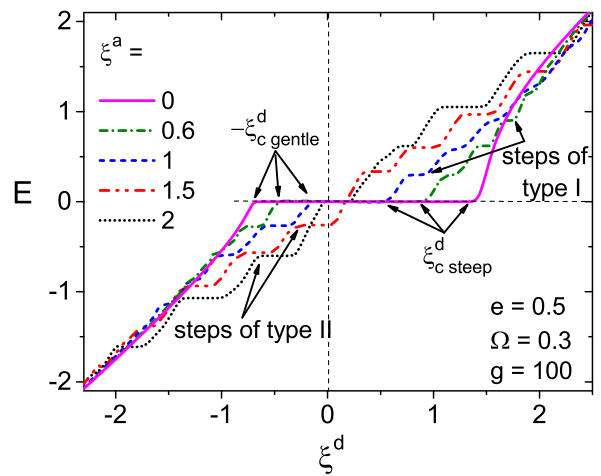}
    \caption{$E(\xi^d)$ by Eq.~\eqref{eCVC} for a series of ac amplitude densities $\xi^a$. Due to the asymmetry of the WPP, the steep-slope $\xi_{c~steep}^{d}$ and gentle-slope $\xi_{c~gentle}^{d}$ depinning dc biases are not equal. The critical bias value corresponds to a vanishing pinning potential barrier, see also Fig. \ref{f4}. Two different types of Shapiro steps are explained in the text.}
   \label{f3}
\end{figure}

\begin{figure}
    \centering
    \includegraphics[width=0.9\linewidth]{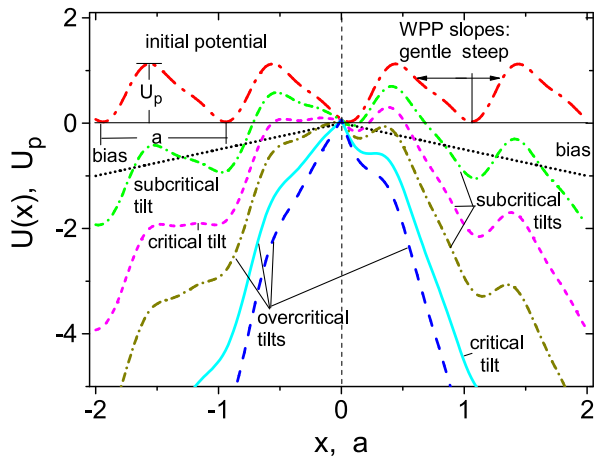}
    \caption{Modification of the effective ratchet WPP $U(x)\equiv U_p(x) - Fx$ with gradual increase of the Lorentz force $F$ in the $x$-direction \cite{Shk14pcm}. The left and right half of the plot correspond to the negative and positive current polarity, respectively. The data are plotted for $e=0.5$ for the ratchet WPP given by Eq.~\eqref{eWPP}. The motive force $F$ is applied against the steep-slope ($F_{p~steep}$ for $x>0$) and the gentle-slope ($F_{p~gentle}$ for  $x<0$) WPP directions. Accordingly, depending on the bias value, in the absence of an ac current and assuming $T=0$ for simplicity, the vortex movement in a tilted ratchet WPP has the following regimes: (i) If $F < F_{p~gentle}$ the vortex is in the localized state. With further increase of the bias value the gentle-slope barrier vanishes, i.\,e., the critical tilt is achieved with respect to the left WPP barrier. (ii) If $F_{p~gentle} < F < F_{p~steep}$, the running mode in the vortex motion appears in the gentle-slope direction of the WPP. The vortex remains in the localized state with respect to the steep-slope direction of the WPP. Further increase of the bias value leads to the vanishing of the right WPP barrier, i.\,e., the steep-slope critical tilt is achieved. (iii) When $F > F_{p~steep}$, the running state in the vortex motion is realized with respect to both, gentle- and steep-slope directions of the WPP.}
   \label{f4}
\end{figure}

Figure~\ref{f2} presents the CVCs calculated for a series of $e$ at the frequency $\Omega = 0.3$ and the overcritical ac amplitude $\xi^a = 1.5$. The meaning of the critical amplitude is explained in Fig.~\ref{f4}. In Fig.~\ref{f3} all CVCs demonstrate Shapiro-like steps regardless of the associated $e$ value. This synchronization effect is generic to systems with a periodic potential subjected to combined dc and ac stimuli. The presence of steps in the CVC for a cosine WPP was reported in \cite{Shk08prb} referring back to an analysis \cite{Van85prb} of the overall shape and positions of steps in the Josephson junction problem, as observed first by Shapiro \cite{Sha63prl}. In contrast to the cosine potential for $e=0$ in Fig. \ref{f2}, several new features appear in the CVCs thanks to the asymmetry of the WPP. First, whereas the regime of viscous flux flow is realized for $e = 0$ over the entire range of dc biases, the CVC for $e = 1$ has a zero plateau on the right-hand branch as long as the critical dc bias for the steep-slope direction of the WPP is not reached. Second, due to the asymmetry of the WPP, the positive (steep-slope) critical current value is greater than the negative (gentle-slope) one $\xi_{c~steep} > |\xi_{c~gentle}|$. Since $\xi^a = 1.5 > |\xi_{c~gentle}| \simeq 0.2$, the threshold in the left branch of the CVC is absent. Finally, the curves for $e = 0.5$ and $1$ demonstrate a \emph{negative} voltage at small \emph{positive} biases. This is a clear indication for the \emph{ratchet reversal effect}, which is most pronounced at $e = 0.5$.

Turning to the influence of the ac amplitude on the shape of the CVCs shown in Fig.~\ref{f3}, the curve for $\xi^a = 0$ should be discussed first. This curve has two unequal-arm branches each having a zero plateau as long as the corresponding critical tilt of the WPP is not reached. In the limit of very strong dc biases $|\xi^d|\rightarrow \infty$ the curve tends to the Ohmic behavior whose slope is determined by the flux-flow resistivity $\rho_f$. If one defines the critical dc bias $\xi^d_c$ as that corresponding to the vanishing WPP barrier in Fig.~\ref{f4}, then it is evident that for the gentle-slope (left) WPP barrier the critical dc density $|-\xi^d_{c~gentle}|$ is less than $\xi^d_{c~steep}$ for the steep-slope (right) WPP barrier. This feature allows for the rectification of ac signals (the diode effect), provided the ac amplitude satisfies the condition $|-\xi_{c~gentle}|\leq\xi^a\leq\xi_{c~steep}$.

With increasing ac amplitude $\xi^a$ several changes in the CVC in Fig.~\ref{f3} should be emphasized. First, for both branches the absolute value of the dc critical density is reduced. Accordingly, $|\xi^d_c(\xi^a)|$ is a decreasing function of the ac drive. The physical reason for this lies in the replacement of the dc critical density by the total (dc+ac) critical amplitude with one additional contribution originating from the asymmetry parameter $e$. This contribution is negative for the steep-slope direction and it is positive for the gentle-slope one \cite{Shk14pcm}.

The second feature is the development of Shapiro-like steps in the CVCs with increasing $\xi^a$. Depending on the relative strength of the ac current amplitude $\xi^a$ with respect to its critical value $\xi^a_c$, two different types of steps in the CVC ensue. The steps of type I appear at $\xi^a > \xi^a_c \equiv 1$ (overcritical ac amplitudes) and distort the CVC like a ripple with a downward shift from the Ohmic line. The stronger $\xi^a$ the broader is the range of $\xi^d$ for these steps to appear. The steps of type II ensue for $\xi^a\geq\xi^a_c$ (subcritical ac amplitudes) and oscillate closely around the Ohmic line. With increasing $\Omega$ the size of the steps of both types increases, whereas their number decreases. In the opposite limiting case of very low frequencies $\Omega\rightarrow0$, i.\,e. in the quasistatic regime, all steps reduce their size along with increasing their number. All steps saturate to the conventional smooth CVC with decreasing ac amplitude.

The third feature consists in the appearance of a negative voltage at small positive dc biases. This ratchet reversal ensues for \emph{overcritical} ac amplitudes with respect to both, the gentle-slope and the steep-slope WPP barriers. The physical reason for the negative voltage to appear is the rectifying effect due to the internal anisotropy of the WPP. An increase of the dc bias superimposes the external tilt-induced asymmetry onto the original intrinsic asymmetry of the WPP, see Fig.~\ref{f4}. This is why the direct net transport starts to prevail over the reversed one.

An examination of the main theoretical predictions for superconducting Nb thin films with an asymmetric washboard pinning nanostructure is reported next.

\section{Experiment}
\label{sExp}
\begin{figure}
    \centering
    \includegraphics[width=0.9\linewidth]{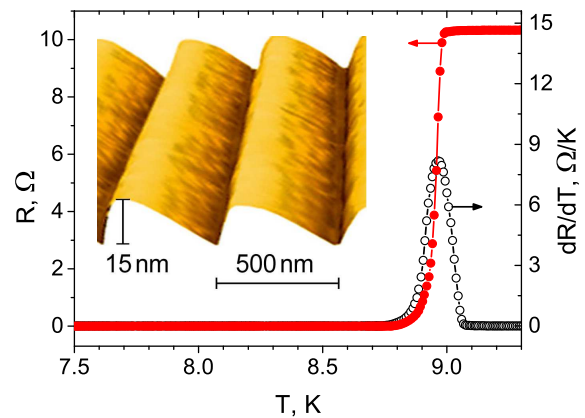}
    \caption{Superconducting transition of the $70$\,nm-thick Nb microstrip patterned with asymmetric grooves by focused ion beam milling. Inset: Atomic force microscope image of the microstrip surface.}
   \label{f5}
\end{figure}

The sample is a $70$\,nm-thick epitaxial Nb (110) film prepared by dc magnetron sputtering onto an a-cut ($11\bar{2}0$) sapphire substrate \cite{Dob12tsf}. The film was pre-patterned by standard photolithography followed by Ar ion-beam etching in order to define a $50\,\Omega$ impedance-matched microstrip with a width of $150\,\mu$m and a length of $500\,\mu$m. A washboard pinning nanolandscape was then fabricated on the microstrip surface by focused ion beam (FIB) milling. The nanopattern is an array of uniaxial straight grooves with a groove-to-groove distance $a$ of $500$\,nm, a depth of $15$\,nm, and a full width at half depth of $200$\,nm. The grooves are aligned parallel to the long side of the microstrip, that is parallel to the transport current direction. They have an asymmetric profile in the cross-section that has been achieved in the patterning process by defining a step-wise increasing number of FIB beam passes to each groove screened as a 5-step ``staircase''. In Fig. \ref{f5}, one sees that due to blurring effects, smooth slopes resulted instead of the ``stairs'' in the nanopattern. It is worth noting that the microstrip width is an integer multiple number ($N=300$) of the nanopattern period and the accuracy of FIB milling in conjunction with the large microstrip dimensions ensure that possible ratchet effects due to the edge barrier asymmetry \cite{Ali09njp} are insignificant to the highest attainable degree. The film is characterized by a critical temperature $T_c$ of $8.94$\,K and by an upper critical field at zero temperature $H_{c2}(0)$ of about $1$\,T, as deduced from fitting the dependence $H_{c2}(T)$ to the phenomenological law $H_{c2}(T) = H_{c2}(0) [1-(T/T_c)^2]$.

Combined microwave and dc electrical resistance measurements were done with magnetic field directed perpendicular to the film surface, using a custom-made sample probe equipped with coaxial cables \cite{Dob15mst}. The microwave signal was generated by an Agilent E5071C vector network analyzer, while the dc signal was added by using two bias-tees mounted at the ports of the analyzer. The CVCs were measured in the fixed-current upsweep mode \cite{Dob15apl,Dob15met}.
The complementary measurements of the microwave power absorbed by vortices \cite{Dob15apl,Dob15met} allowed us to deduce the coordinate dependence of the mean pinning potential in a series of samples with different grooves' asymmetry degrees, using the procedure outlined in \cite{Shk12inb,Shk13ltp}. Accordingly, for the sample used in this work the coordinate dependence of the pinning potential has been approximated by Eq. \eqref{eWPP} with $e =0.56$ which is very close to $e = 0.5$ used for the simulation results presented in Sec. \ref{sTheor}.
\begin{figure*}
    \centering
    \includegraphics[width=0.9\textwidth]{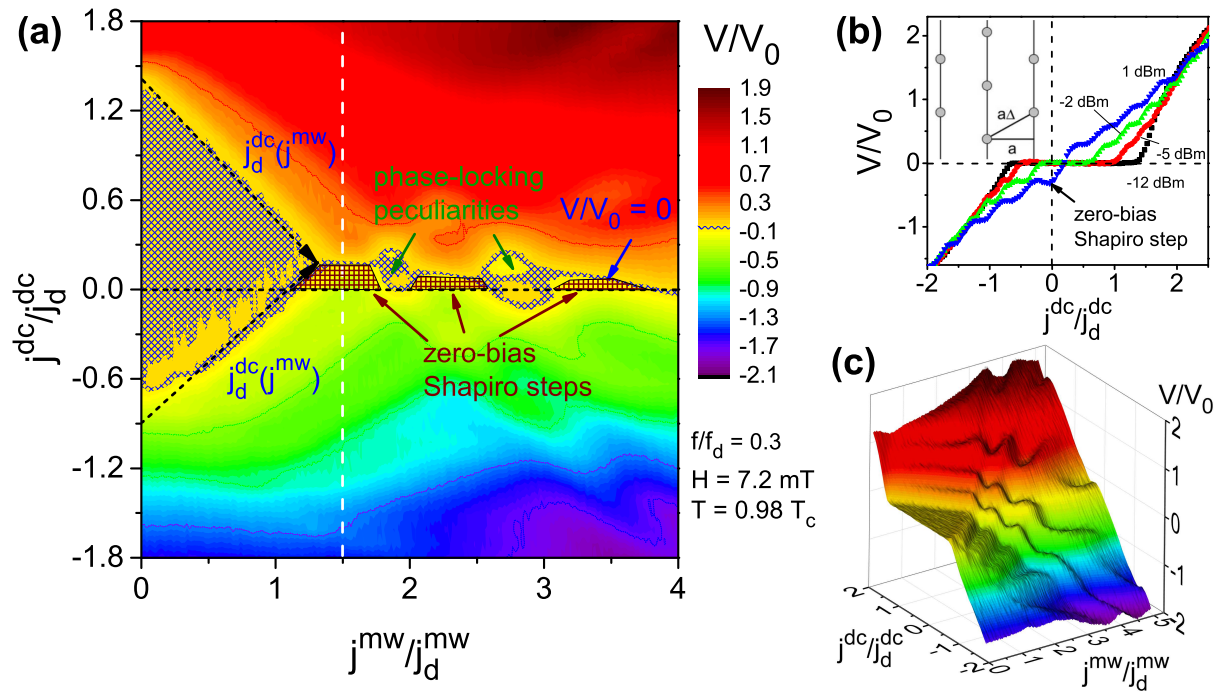}
    \caption{(a) Electrical dc voltage as a function of the dc value and the microwave amplitude.
     The dashed vertical line corresponds to the ac amplitude for which the zero-bias Shapiro step is marked in (b).
    (b) CVC for a series of microwave amplitudes expressed in terms of the nominal power in dBm. The arrangement of vortices with respect to the nanostructure is shown in the inset. (c) The same as (a), but in 3D representation. In all panels $T = 0.98T_c$ and $H = 7.2$\,mT, the voltage is normalized by the first Shapiro step voltage, while the currents by the corresponding depinning values.}
   \label{f6}
\end{figure*}

Figure \ref{f6} displays the dc electrical voltage as a function of the normalized dc density and the ac amplitude for the ac frequency $f = 906$\,MHz. The measurements are done at the temperature $T = 0.98T_c$ and the fundamental matching field $H = 7.2$\,mT. The arrangement of vortices at $7.2$\,mT with respect to the pinning nanolandscape is shown in the inset of Fig. \ref{f6}(b) for the assumed triangular vortex lattice with lattice parameter $a_\bigtriangleup = (2\Phi_0/B\sqrt{3})^{1/2}$ and the matching condition $a_\bigtriangleup = 2a/\sqrt{3}$. In the absence of microwave excitation, the CVC in Fig. \ref{f6}(b) demonstrates two different absolute values of the depinning current density $j_d^+ > |j_d^-|$ for the positive and the negative branch. The depinning current density $j_d$ is determined by the $0.1\,\mu$V voltage criterion, while we use the mean-square parameter $j_d=\sqrt{j_d^+ j_d^-}$ for the presentation of the data in dimensionless form allowing for a direct comparison of experiment with theory. At $T=0.3T_c$ and $H=7.2$\,mT the depinning current densities for the gentle-slope and the strong-slope directions of the nanostructure amount to $|j_d^-| = 0.52$\,MA/cm$^2$ and $j_d^+ = 1.25$\,MA/cm$^2$, respectively. Accordingly, this yields the mean-square depinning current $80$\,mA.

The addition of the microwave stimulus leads to the appearance of Shapiro steps in the CVC. The steps occur at voltages \cite{Fio71prl}
\begin{equation}
    \label{eCVCexp}
    V = n V_0 \equiv nN\Phi_0 f,
\end{equation}
where $n$ is an integer, $N$ is the number of vortex rows between the voltage leads, $f$ is the microwave frequency, and $\Phi_0 = 2.07\times10^{-15}$~Vs is the magnetic flux quantum. The steps in the CVCs arise when one or a multiple of the hopping period of Abrikosov vortices coincides with the period of the ac drive. One can distinguish up to five lowest-order Shapiro steps and from the step voltage one can infer $N=860$-$870$. This is similar to the Nb films with symmetric grooves dealt with previously \cite{Dob15snm}. Given the geometrical dimensions of the microstrip and the fundamental matching field configuration for a triangular vortex lattice [see the inset to Fig.~\ref{f6}(b)], the expected number of vortex rows between the voltage leads is equal to $866$. The fact that the number of vortex rows deduced from fitting the experimental data to Eq.~(\ref{eCVCexp}) is very close to $866$ suggests that all vortices move coherently. This strongly coherent motion is caused by both, the high periodicity of the nanogroove array and a relatively weak contribution of the background isotropic pinning due to structural imperfectness as compared to the dominating strong pinning owing to the nanopatterning. This conclusion is in line with our previous observations \cite{Dob12njp,Dob15snm} that the focused ion beam-milled grooves provide a strong pinning potential for vortices forced to move across them. When tuning the field value away from the matching configuration the steps disappear.

Turning to the general description of the contour plot in Fig. \ref{f6}(a), several features should be noted. First, the dc depinning current value nearly linearly decreases with increasing microwave amplitude, i.\,e. the experimental data confirm the theoretical prediction that the microwave current density contributes to the depinning of vortices. Second, at $j^{mw}/j^{mw}_d \approx 1.25$ the dc depinning values turn out to be zero and this leads to the effective ``symmetrization'' of the CVC, but with the origin shifted by $j^{dc}/j^{dc}_d\approx0.25$ towards the positive bias values. The theory suggests that at this dc bias the internal asymmetry of the pinning potential is effectively compensated by the extrinsic asymmetry of the WPP caused by the dc bias. This allows one to interpret this dc value as that characterizing the \emph{loading capability} of the ratchet \cite{Knu12pre}. Third, at relatively small \emph{positive} dc values $j^{dc}/j^{dc}_d \leq 0.2$ for the microwave amplitudes $1 \lesssim j^{mw}/j^{mw}_d \lesssim 4$ a small \emph{negative} voltage $|V|/V_0\lesssim 0.5$ is revealed. With further increasing dc value the voltage tends to zero and changes its sign, that is the ratchet reversal is observed. Finally, at some microwave amplitudes an enhancement of the dc voltage is observed, as designated by ``phase-locking peculiarities'' in Fig. \ref{f6}(a). Similar peculiarities were theoretically predicted earlier for a symmetric \cite{Shk11prb} and an asymmetric \cite{Shk14pcm} WPPs so that we believe that the features observed experimentally correspond to the three lowest-order mode-locking fringes discussed in \cite{Shk14pcm}.

\section{Conclusion}
In the presence of combined dc and microwave drives the vortex dynamics in asymmetric pinning potentials exhibits several dynamical regimes which are determined not only by the magnitude of the applied current, but also by its polarity. These different regimes in the vortex dynamics have been analyzed in the framework of a stochastic single-vortex model on the basis of the Langevin equation. This allows for a direct comparison of the coherent vortex dynamics of the vortex ensemble in Nb films with asymmetric grooves at the fundamental matching field with the theoretical predictions derived in the single-vortex approach. This early experiment has revealed the main features of the current-voltage curves predicted theoretically, namely:
\begin{itemize}
  \item The CVC of a superconductor with an asymmetric WPP is characterized by two different depinning current values in the positive and the negative branch.
  \item The dc depinning current density values for both dc polarities decrease linearly with increasing microwave amplitude. The microwave amplitude corresponding to the intersection of the $j^{dc}_d(j^{mw})$-dependences for both dc polarities allows one to infer the dc value at which the internal asymmetry of the pinning potential is compensated by its external asymmetry induced by the tilt due to the dc bias.
  \item A switching between the positive and the negative ratchet effect (ratchet reversal) takes place depending on the balance between the fixed internal and the dc-bias tunable asymmetry of the pinning potential.
  \item A negative voltage at zero dc bias becomes apparent as zero-bias Shapiro steps which can be explained as a synchronization effect in the current-voltage curve shifted from its origin due to the asymmetry of the pinning potential.
\end{itemize}

\vspace{3mm}
This work was financially supported by the German Research Foundation (DFG) through grant DO 1511 and conducted within the framework of the NanoSC-COST Action MP1201 of the European Cooperation in Science and Technology. This research has received funding from the European UnionТs Horizon 2020 research and innovation program under Marie Sklodowska-Curie Grant Agreement No. 644348 (MagIC).

%\bibliography{D:/bibliobase/fluxonics}
%\bibliographystyle{spphys}
%\bibliographystyle{elsarticle-num}

\end{document}